\pdfoutput=1
\documentclass[twocolumn,english,aps]{revtex4}
\usepackage[T1]{fontenc}
\usepackage[latin9]{inputenc}
\usepackage{color}
\usepackage{babel}

\usepackage{amsmath}
\usepackage{graphicx}
\usepackage{amssymb}
\usepackage[unicode=true,pdfusetitle,
 bookmarks=true,bookmarksnumbered=false,bookmarksopen=false,
 breaklinks=false,pdfborder={0 0 1},backref=false,colorlinks=false]
 {hyperref}

\makeatletter
\@ifundefined{textcolor}{}
{%
 \definecolor{BLACK}{gray}{0}
 \definecolor{WHITE}{gray}{1}
 \definecolor{RED}{rgb}{1,0,0}
 \definecolor{GREEN}{rgb}{0,1,0}
 \definecolor{BLUE}{rgb}{0,0,1}
 \definecolor{CYAN}{cmyk}{1,0,0,0}
 \definecolor{MAGENTA}{cmyk}{0,1,0,0}
 \definecolor{YELLOW}{cmyk}{0,0,1,0}
 }

\makeatother

\makeatother

\begin{document}
\global\long\global\long\def\ket#1{|#1\rangle}

\global\long\global\long\def\bra#1{\langle#1|}

\global\long\global\long\def\proj#1{|#1\rangle\langle#1|}

\global\long\global\long\def\ketbra#1#2{|#1\rangle\langle#2|}

\global\long\global\long\def\braket#1#2{\langle#1|#2\rangle}

\title{Practical learning method for multi-scale entangled states}

\author{Olivier Landon-Cardinal}

\affiliation{D\'epartement de Physique, Universit\'e de Sherbrooke, Sherbrooke,
Qu\'ebec,
J1K 2R1, Canada}

\author{David Poulin}

\affiliation{D\'epartement de Physique, Universit\'e de Sherbrooke, Sherbrooke,
Qu\'ebec,
J1K 2R1, Canada}
\begin{abstract}
We describe a method for reconstructing multi-scale entangled states
from a small number of efficiently-implementable measurements and
fast post-processing. The method only requires single-particle measurements
and the total number of measurements is polynomial in the number of
particles. Data post-processing for state reconstruction uses standard
tools, namely matrix diagonalisation and conjugate gradient method,
and scales polynomially with the number of particles. Our method prevents
the build-up of errors from both numerical and experimental imperfections.
\end{abstract}
\maketitle

\section{Introduction}

Quantum state tomography \cite{Vogel1989} is a method to learn a
quantum state from measurements performed on many identically prepared
systems. This task is crucial not only to assess the degree of control
exhibited during the preparation and transformation of quantum states,
but also in comparing theoretical predictions to real-life systems.
For instance, numerical methods are used to compute the ground states
or thermal states of model quantum systems. Quantum state tomography
could be used to check that the experimental state corresponds to
the predicted one, thus providing an essential link between theory
and experiments. For example, one could in principle use tomography
to settle the question \cite{DGZ10} of which states correctly describe
the quantum Hall fluid at various filling parameters.

In practice however, the state of $n$ particles is described by a
number of parameters that scales exponentially with $n$. Therefore,
tomography requires an exponential number of identically prepared
systems on which to perform exponentially many measurements needed
to span a basis of observables that completely characterizes the state.
Furthermore, solving the inference problem to determine the quantum
state that is compatible with all these measurement outcomes requires
an exponential amount of classical post-processing. These factors
limit tomography to at most a few tens of particles.

While this is unavoidable for a generic state, many states encountered
in nature have special properties that could be exploited to simplify
the task of tomography. In fact, the overwhelming majority of tomographic
experiments performed to date \cite{Haffner2005,Barreiro2010,DiCarlo2009,DiCarlo2010,Filipp2009,Mikami2005,Resch2005}
were used to learn state described with only a few parameters. Such
variational states---family of states specified with only a few parameters---are
widespread in many-body physics because they are tailored for numerical
calculations and can predict many phenomena observed in nature (Kondo
effect, superconductivity, fractional statistics, etc). One example,
familiar to the quantum information community, is matrix product states
(MPS) \cite{Affleck1987,Fannes1992,Vidal2003,Vidal2004} that are
at the heart of the density matrix renormalisation group (DMRG) numerical
method, suitable for the description of one-dimensional quantum systems
with finite correlation length \cite{VC06}. 

Recently, we and others have demonstrated \cite{CPF+10} that tomography
can be performed efficiently on MPS, \textit{i.e.}, such states can
be learned from a small number of simple measurements and efficient
classical post-processing. Here, we take this result one step further
and demonstrate that it is possible to efficiently learn the states
associated to the multi-scale renormalization ansatz (MERA) introduced
by Vidal \cite{Vidal2008}, for which efficient numerical algorithms
to minimize the energy of local Hamiltonians exist \cite{Evenbly2009}.
As opposed to MPS, these MERA states are not restricted to one dimension
and can describe systems with long-range correlation. This last distinction
is important because one of the most interesting phenomena in physics,
quantum phase transitions, leads to a diverging correlation length
and are therefore not suitably described by MPS. In contrast, MERA
have been successfully used to study numerous many-body models, such
as the critical Ising model in 1D \cite{Evenbly2009} and 2D \cite{Cincio2008},
and can also accurately describe systems with topological order \cite{Aguado2008,Evenbly2009a,Montangero2009,Pfeifer2009,Konig2010,Pfeifer2010}.

In this work, we present two related methods to learn the one-dimensional
MERA description of a state using tomographic data obtained from local
measurements performed on several copies of the states. Our learning
methods for MERA are based on the identification of the unitary gates
in the quantum circuit that outputs the MERA state. In that regard,
this Article is a continuation of our work on MPS and is reminiscent
of early methods to numerically optimise MERA tensors \cite{Vidal08}.
However, going from MPS to MERA is non-trivial because MERA exhibits
a spatial arrangement of gates that is more elaborate. Since MERA
is a powerful numerical tool, our learning method bridges the gap
between numerical simulations and experiments by allowing the direct
comparison of numerical predictions to experimental states.

The first method we present requires unitary control of the system
and the ability to perform tomography on blocks of a few particles,
which can be realized using the correlations between single-particle
measurements. Crucially, the size of those blocks does not depend
on the total size of the system, making it a \emph{scalable} method.
\textcolor{black}{In an experiment, one cannot know beforehand if
the state in the lab is a MERA. However, our method contains a
built-in certification procedure from which one can assess the proper
functioning of the method as the experiments are performed and conclusively
determine if the state is well described by the MERA. The second method
builds on the first one, but completely circumvents the need for unitary
control. Thus, this MERA learning method can be implemented with existing
technologies. The drawback of this simplified method is that it does
not come with a built-in certification procedure. Certification in
this case can be realized using the Monte Carlo scheme \cite{SLP11},
which requires the same experimental toolbox.}

The rest of the paper is organised as follows. We first present the
proposed method for MERA learning in section \ref{sec:MERA-tomography}.
Subsection \ref{sub:Identifying-the-disentanglers} explains how to
identify the disentanglers. We start by deriving a necessary condition
for the existence of suitable disentangler and then turn this criterion
into a heuristic objective function that we minimize numerically over
unitary space. In subsection \ref{sub:Error-analysis}, we carefully
analyze the buildup of errors in our procedure and show that errors
only accumulate linearly with the size of the system. In subsection
\ref{sub:Numerical-performance}, we present numerical benchmarks
of our tomography method. In subsection, \ref{sub:Trading-unitary-control},
we present the simplified method by demonstrating that it is not necessary
to apply the disentanglers to the experimental state since we can
simulate the effect of those disentanglers numerically, albeit at
the cost of more repeated measurements and a slightly worse error
scaling (analyzed in appendix \ref{sec:Error-analysis-no-post-select}).
In section \ref{sec:Discussion}, we discuss potential issues for
our numerical scheme and suggest modifications to prevent them. Finally,
we present in appendix \ref{sec:Contraction} a tool to contract two
different MERA states, which allows for the efficient comparison of
a MERA whose parameters have been identified experimentally using
our method to a predicted theoretical MERA state.

\section{\label{sec:MERA-tomography}MERA learning}

\subsection{Identifying the disentanglers\label{sub:Identifying-the-disentanglers}}

MERA states can be described as the output of a quantum circuit \cite{Vidal2008}
whose structure is represented on Fig. \ref{fig:optimal-disentangler}
(as seen with inputs on the top and output at the bottom). For simplicity,
we will focus on a one dimensional binary MERA circuit for qubits,
but our method generalizes to all 1D MERA states, \textit{i.e.}, particles
could have more internal states thus accounting for a larger MERA
refinement parameter $\chi$ and isometries could renormalize several
particles to one effective particle. The circuit contains three classes
of unitaries. Disentanglers (represented as $\square$) are two-qubit
unitary gates; isometries (represented as $\triangle$) are also two-qubit
gates but with with one input qubit always in the $\ket 0$ state;
the top tensor (represented as $\bigcirc$) is a special case of isometry
that takes as input two qubits in the $\ket{00}$ state. Each renormalisation
layer is made of a row of disentanglers and a row of isometries. Disentanglers
remove the short-scale entanglement between adjacent blocks of two
qubits while isometries renormalise each pair of qubits to a single
qubit. Each renormalisation layer performs theses operations on a
different lengthscale. The quantum circuit thus mirrors the renormalisation
procedure that underlies the MERA.

Learning a MERA amounts to identifying the various gates in that circuit,
which turns the experimental state into the all $\ket 0$ state. The
intuitive idea behind our scheme is to proceed by varying the isometries
and disentanglers until the {}``ancillary'' qubits reach the state
$\ket 0$ for each row of isometries. We will exploit this feature
to numerically determine each disentangler.

\subsubsection{Necessary condition for disentangler}

Consider the $n$ qubits at the lowest layer of the MERA. %
\begin{figure}
\begin{centering}
\includegraphics[width=1\columnwidth]{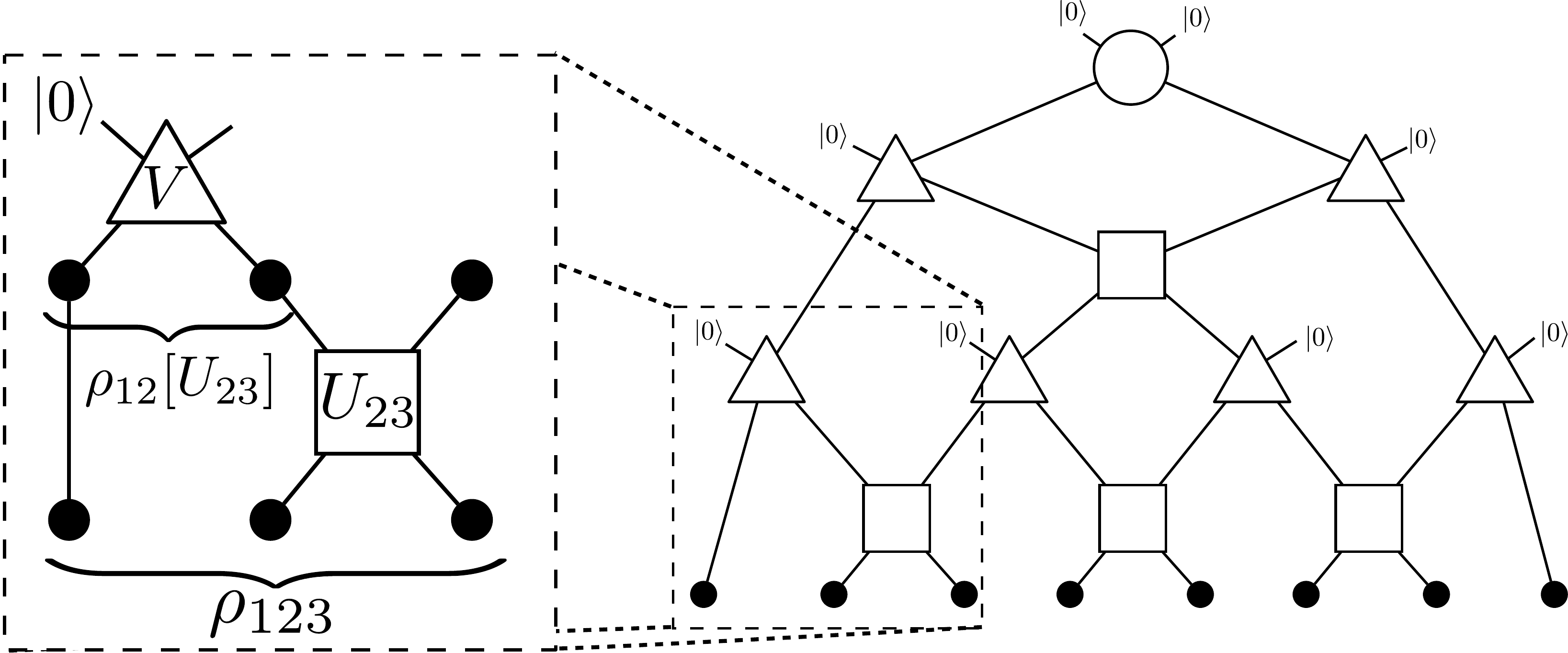} 
\par\end{centering}

\caption{\label{fig:optimal-disentangler}The optimal disentangler $\tilde{U}$
can be computed from the tomographic estimation of the density matrix
$\rho_{123}$ on the first three qubits. Once applied, the resulting
state $\tilde{\rho}_{12}[\tilde{U}]$ is very close to a rank 2 matrix.
Thus, there exist a unitary $V$ such that $\tilde{\rho}_{12}[\tilde{U}]$
can be rotated such that the first qubit is almost in the state $\ket 0$.}
\end{figure}
 Let $\rho_{123}$ be the reduced density matrix on the three first
qubits (see Fig. \ref{fig:optimal-disentangler}). If the state is
exactly a MERA, there exists a unitary $U_{23}$ acting on qubits
2 and 3 (see left of Fig. \ref{fig:optimal-disentangler}) such that
applying this unitary and tracing out the 3rd qubit yields a density
matrix \begin{equation}
\rho_{12}\left[U_{23}\right]=\mbox{Tr}_{3}\left[\left(\mathbb{I}_{1}\otimes U_{23}\right)\rho_{123}\left(\mathbb{I}_{1}\otimes U_{23}^{\dagger}\right)\right]\label{eq:StateAfterDisentangler}\end{equation}
\emph{of rank at most 2}. Indeed, if the rank was strictly greater
than 2, it would be impossible for the isometry $V$ (see left of
Fig. \ref{fig:optimal-disentangler}) to map the density matrix $\rho_{12}\left[U_{23}\right]$
to a state with one of the qubit in the state $\ket 0$ because the
dimension of the space $\ket 0\otimes\mathbb{C}^{2}$ would be strictly
smaller than the dimension of the support of the density matrix. Hence,
we have the necessary criterion\begin{equation}
\exists\tilde{U}_{23}\quad\rho_{12}\left[\tilde{U}_{23}\right]\mbox{ has rank less or equal than 2\ensuremath{.}}\end{equation}

To find a unitary that fulfills this criterion, it is necessary to
know the state $\rho_{123}$, and this can be achieved by brute-force
tomography on these three qubits. Once the original state on the three
qubits is known, one has to perform a search over the space of unitaries
to find a suitable disentangler. To do this, we will define an objective
function to minimise numerically. 

Once this optimal unitary operator $\tilde{U}$ has been found numerically,
it is necessary to consider how it modifies the quantum state before
learning the other elements of the circuit. One obvious way to do
so is to apply the unitary transformation to the experimental state
and continue the procedure on the transformed state. This amounts
to undoing the circuit, and should in the end map the experimental
state to the all $\ket 0$ state. For simplicity, we will first present
our scheme assuming that the state is transformed at every step this
way. Of course, such unitary control increases the complexity of the
scheme and could be out of the reach of current technologies. However,
in section \ref{sub:Trading-unitary-control}, we will explain how
this unitary transformation can be circumvented at the cost of a slight
increase in the number of measurements.

After the optimal disentangler $\tilde{U}$ has been applied to the
state, we need to identify the unitary $V$ that rotates the density
matrix on the first two qubits such that the first qubit is brought
to the $\ket 0$ state, c.f. Fig \ref{fig:optimal-disentangler} left.
This does not require any additional tomographic estimate since we
already know the descriptions of the state on the three first qubits
and the disentangler. We can thus compute the state on the first two
qubits $\rho_{12}[\tilde{U}]$ and diagonalise it to obtain the eigenvectors
corresponding to its two non-zero eigenvalues. The unitary $V$ is
chosen to map those two eigenvectors to the space $\ket 0\otimes\mathbb{C}^{2}$,
i.e., $V$ rotates the qubits such that the support of the density
matrix is mapped to a space where the first qubit is in the $\ket 0$
state.

All other disentanglers of this layer can be found by recursing the
above procedure. Once a disentangler has been identified,
it is physically applied to the system and brute-force tomogrography
is performed on the next block of three qubits. 

\begin{figure}
\begin{centering}
\includegraphics[width=1\columnwidth]{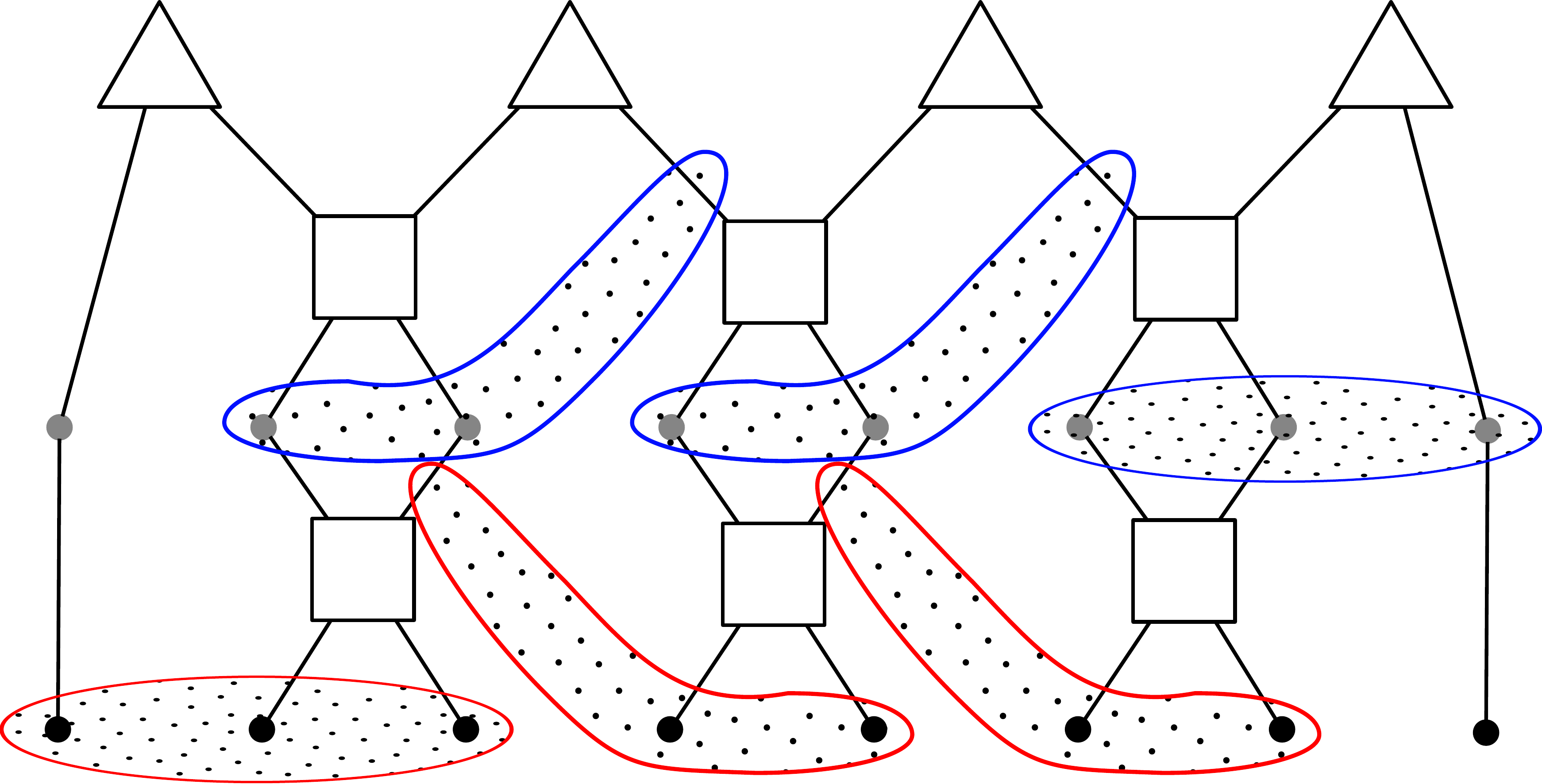}
\par\end{centering}

\caption{\label{fig:multiple-sweeps-schema}Identification of the disentanglers
using two successive sweeps of the chain. Dotted regions cover particles
on which brute-force tomography is performed. The first sweep (red
dotted regions) finds unitaries starting from the left end of the
chain. Those unitaries will be used as initial guesses for the second
sweep (blue striped regions) that starts from the right end of the
chain.}
\end{figure}

Notice that for the last block, a single unitary is responsible for
minimising the rank of two density matrices, $\rho_{n-3,n-2}$ and
$\rho_{n-1,n}$. One possible way to handle
this is to get a tomographic estimate of the state on the last four
qubits and to try to minimise the rank of both reduced matrices. Another
way, for which we have opted in our numerical simulations, is to perform
multiple sweeps over the layer. For instance, the disentanglers will
first be identified from left to right and then the next sweep will
be performed from right to left, using the disentanglers found in
the first sweep as initial guesses in the space of unitaries (see
Fig. \ref{fig:multiple-sweeps-schema}). %
 The number of sweeps can be increased for better accuracy but each
additional sweep requires to extract the tomographic estimates again.
Multiple sweeps would also allow to apply our method to MERA states
with periodic boundary conditions in 1D and could be useful for 2D-MERA
states. While this would be an interesting continuation of our work,
we focus on 1D-MERA for the rest of the article.

\subsubsection{Heuristic objective functions}

One of the steps in our protocol consists in identifying the unitary
$\tilde{U}$ that minimizes the rank of $\rho_{12}[U]$, c.f. eq.
\eqref{eq:StateAfterDisentangler}. There are many distinct ways this
can be done and in this section, we present a practical heuristic
to accomplish this task. Minimising the rank of the density matrix
$\rho_{12}\left[U\right]$ is not a suitable numerical task because,
even if the experimental state is an exact MERA, the inferred density
matrix will typically have full rank due to machine precision and
the imperfect tomographic estimation of $\rho_{123}$. Thus, we turn
the problem of finding $\tilde{U}_{23}$ into an optimization problem
by considering the eigendecomposition of $\rho_{12}\left[U\right]=\sum_{k}\lambda_{k}\proj{\psi_{k}}$
where the eigenvalues are sorted in decreasing order $\lambda_{1}\geq\lambda_{2}\geq\lambda_{3}\geq\lambda_{4}$.
If $\rho_{12}[U]$ has most of its support on a two-dimensional space,
it will have two small eigenvalues that are typically non-zero due
to imperfections. We thus consider the objective function\begin{equation}
f\left(U,\rho_{123}\right)=\sum_{k>2}\lambda_{k}\label{eq:Objective-function}\end{equation}
 and we perform a minimisation over the space of unitaries to determine
the optimal unitary $\tilde{U}$. This objective function has a well-defined
operational meaning -- it is the probability of measuring the disentangled
qubit in the $\ket 1$ state after the isometry $V$ has been applied.
We will see in section \ref{sub:Error-analysis} that this property
can be used to certify the distance between the experimental and the
reconstructed states.

\textcolor{black}{Another way to think about this objective function
is to consider the characteristic polynomial $P[X]$ of $\rho_{12}\left[U\right]$
which is of the form\begin{equation}
P[X]=X^{4}-X^{3}+aX^{2}-bX+c\end{equation}
where the coefficients $a$, $b$ and $c$ are positive since they
correspond to the sum of product of the positive eigenvalues of the
density matrix. In particular, coefficient $b$ is the sum of all
products of three eigenvalues, }\textit{\textcolor{black}{i.e.}}\textcolor{black}{,
$b=\lambda_{1}\lambda_{2}\lambda_{3}+\lambda_{1}\lambda_{2}\lambda_{4}+\lambda_{1}\lambda_{3}\lambda_{4}+\lambda_{2}\lambda_{3}\lambda_{4}$.
In order for the rank of the density matrix to be 2, it is sufficient
for all 4 products of three eigenvalues to vanish, }\textit{\textcolor{black}{i.e}}\textcolor{black}{,
\begin{eqnarray}
\rho_{12}\left[U\right]\mbox{ of rank less than 2} & \Longleftrightarrow & b=0\mbox{.}\end{eqnarray}
Thus, another suitable objective function is the positive coefficient
$b$, which is a polynomial in the entries of $\rho_{12}\left[U\right]$.
Indeed, using Bocher formula, coefficient $b$ can be expressed as
$6b=1-3\mbox{Tr}A^{2}+2\mbox{Tr}A^{3}$ where $A=\rho_{12}\left[U\right]$.
Thus, $b$ is a well-behaved function with respect to the density
matrix. Note also that $b$ can be expressed without diagonalising
the density matrix $\rho_{12}[U]$. We will focus on minimising \eqref{eq:Objective-function}
in all subsequent numerical discussion and results.}

\subsubsection{Numerical minimisation over unitary space}

Minimisations of \eqref{eq:Objective-function} is performed using
a conjugate gradient method. We first have to account for the fact
that the unitary manifold is not a vector space. To get around this
problem, we go to the Hermitian space by writing any unitary $U$
as the result of a Hamiltonian evolution, i.e., there exists a Hermitian
matrix $H$ such that $U=e^{iH}$. It is then possible to use the
standard conjugate gradient method. Let us sketch the algorithm in
more details.

First, we select a unitary $U_{0}$ either at random or from an initial
guess (provided for instance by a previous sweep).  We will search
the unitary space by generating a sequence of unitaries $\left\{ U_{k}\right\} $.
At the $k$\textsuperscript{th} step of the minimisation, the algorithm
is the following.
\begin{enumerate}
\item We center the unitary space at point $U_{k-1}$ by defining $\rho_{k}=(\mathbb{I}\otimes U_{k-1})\rho_{k-1}(\mathbb{I}\otimes U_{k-1})^{\dagger}$.
\item We compute the gradient $G^{(k)}$ by parametrizing the Hamiltonian
$H$ on 2 qubits by its decomposition on the Pauli group $H=\sum_{\mu\nu}h_{\mu\nu}\sigma_{\mu}\otimes\sigma_{\nu}$
where $\sigma_{\mu}\in\{\mathbb{I},\sigma_{x},\sigma_{y},\sigma_{z}\}$
is a Pauli matrix. We successively evaluate the component of the gradient
$G^{(k)}$ in the direction $(\mu,\,\nu)$ by looking at the effect
of the test unitary $U_{\mu,\nu}=\mathbb{I}+i\epsilon\sigma_{\mu}\otimes\sigma_{\nu}$
on the objective function, \textit{i.e.}, $G_{\mu,\,\nu}^{(k)}=\frac{f(U_{\mu,\,\nu},\,\rho_{k})-f(\mathbb{I},\,\rho_{k})}{\epsilon}$
where $\epsilon$ is a small number.
\item Instead of following the gradient, which would generally undo some
of the minimization performed in the previous steps, we use a conjugate
gradient method where the new direction of search $\tilde{G}^{(k)}$
is optimized by taking into account the direction used in the previous
step $\tilde{G}^{(k-1)}$ through the Polak-Ribi�re formula. More
precisely, $\tilde{G}^{(k)}=G^{(k)}+\beta\tilde{G}^{(k-1)}$ in which
the real parameter $\beta$ is defined as $\beta=\max\left(0,\,\frac{G^{(k)}\cdot\left(\tilde{G}^{(k-1)}-G^{(k)}\right)}{\tilde{G}^{(k-1)}\cdot\tilde{G}^{(k-1)}}\right)$.
\item We perform a line search along the direction $\tilde{G}^{(k)}$ by
considering the family of unitaries $\exp\left(-it\sum_{\mu,\nu}\tilde{G}_{\mu,\nu}\sigma_{\mu}\otimes\sigma_{\nu}\right)$
and optimizing the parameter $t$ to find $t_{opt}$. We then define
\begin{equation}
U_{k}=\exp\left(-it_{opt}\sum_{\mu,\nu}\tilde{G}_{\mu,\nu}\sigma_{\mu}\otimes\sigma_{\nu}\right)U_{k-1}\end{equation}
which ends the $k$\textsuperscript{th} iteration. 
\end{enumerate}
We iterate until the objective function is close enough to zero or
that improvement has stopped.

This method is \emph{heuristic} since the objective functions present
no characteristic that would ensure the convergence of the conjugate
gradient method. In particular, our search over unitary space depends
on the starting point, \textit{i.e.}, the unitary chosen in the first
iteration. Indeed, some starting points will lead the heuristic to
a local minima where it will get stuck. In order to avoid this phenomenon,
we can repeat the overall search by picking at random (according to
the unitary Haar measure) different initial points which lead to potentially
different minima and keep the smallest of those minima, which we expect
to be the global minimum. In any case, this is a minimization problem
over a spapce of \emph{constant} dimension, so the method used to
solve it does not affect the scaling with the number of particles
$n$. Ultimately, we can always use a finite mesh over the unitary
space and use brute-force search. Nevertheless, we found numerically
that this heuristic works well. 

It is also possible that a choice of unitary that is optimal locally,
in the sense that it minimizes \eqref{eq:Objective-function}, is
not optimal globally as it might lead to a state for which it is impossible
to find a disentangler obeying \eqref{eq:Objective-function} elsewhere
in the circuit. This is a phenomenon that is more likely to occur
when the minimum is degenerate, \textit{i.e.}, there exists several
distinct (modulo gauge) exact disentanglers for the state. However,
we have performed numerical experiments on randomly generated MERA
states as well as physically motivated states and found that the conjugate
gradient performs well (see section \ref{sub:Numerical-performance}).

\subsection{\label{sub:Error-analysis}Error analysis}

In practice, due to numerical and experimental imperfections, the disentangled
qubits will not be exactly in the $\ket 0$ state,
but merely close to it. This situation arises from the conjunction
of three causes : \textit{i)} the experimental state of the system
is not exactly a MERA, but merely close to one, \textit{ii)} the tomographic
estimate of the density matrices on blocks of three qubits are slightly
inaccurate due to noisy measurements and experimental finite precision,
\textit{iii)} the numerical minimization did not find the exact minimum.

\subsubsection{Isolating each elementary steps to prevent error buildup}

Our error analysis will show that the buildup of errors is linear
in the number of disentanglers of the MERA circuit which is itself
linearly proportional to the number of particles in the experimental
state. Essentially, the distance between the reconstructed state and
the experimental state is the sum of the error made at each elementary
step when estimating a disentangler and an isometry. Fortunately,
the error made at each elementary step does not depend on the errors
made at previous steps. The key to isolate each step from the others
is to measure the qubit that should have been disentangled in the
computational basis. With high probability the qubit will be found
in the $\ket 0$ state. While the probability of measuring the $\ket 0$
outcome depends on previous errors, the post-selected state is now
free from previous errors. The interest of this postselection is two-fold.
First, it forbids errors in previous steps to contaminate the state
and amplify the error made at the current step, thus limiting the
error propagation. Second, by accumulating statistics on this measurement,
we can estimate the probability of the all-0 outcome and use it to
bound the distance of the reconstructed state to the actual state
in the lab. Therefore, our procedure comes with a \emph{built-in certification
process}. Let us now describe the error analysis in more details.

\subsubsection{Error at each elementary step}

Recall the notation of Fig. \ref{fig:optimal-disentangler}. Due to
numerical and experimental imperfections, the state on qubits 1, 2
and 3 after applying the disentangler $\tilde{U}_{1}$ and the isometry
$V_{1}$ is not exactly in the $\ket 0\otimes\mathbb{C}^{2(n-1)}$
subspace but contains a small component orthogonal to that space.
Thus, it has the form \begin{equation}
V_{1}\tilde{U}_{1}\ket{\psi}=\frac{\ket 0\ket{\eta_{1}}+\ket{e_{1}}}{\sqrt{1+\braket{e_{1}}{e_{1}}}}\end{equation}
where $\ket{\eta_{1}}$ is the normalised pure state on qubits 2 to
$n$ if qubit 1 had been completely disentangled from the chain and
$\ket{e_{1}}$ is some \emph{subnormalized} vector supported on the
subspace $\ket 1\otimes\mathbb{C}^{2(n-1)}$ . The isometry $V_{1}$
is chosen to minimize the norm of $\ket{e_{1}}$, i.e., to minimize
$\epsilon_{1}\equiv\braket{e_{1}}{e_{1}}$. 

Further along the layer, the state after applying $k$ disentanglers
and $k$ isometries will be of the form \begin{equation}
V_{k}\tilde{U}_{k}\dots V_{1}\tilde{U}_{1}\ket{\psi}=\frac{\ket 0^{\otimes k}\ket{\eta_{k}}+\ket{e_{k}^{cm}}}{\sqrt{1+\epsilon_{k}^{cm}}}\label{eq:state-with-error}\end{equation}
where the first term $\ket 0^{\otimes k}\ket{\eta_{k}}$ is the normalised
state had the $k$ qubits in position 1, 3$\dots2k-3$ been completely
disentangled from the chain and $\ket{e_{k}^{cm}}$ is the accumulated
error vector orthogonal to the space where those $k$ qubits are in
the $\ket 0^{\otimes k}$ state. In order to find the optimal disentangler
and isometry, we measure the last disentangled qubit in the computational
basis and post-select on the $\ket 0$ outcome, which occurs with
probability $(1+\epsilon_{k}^{cm})^{-1}$. We then perform brute force
tomography and identify numerically the disentangler and the isometry
that minimimizes the norm of the error vector $\ket{e_{k+1}}$ such
that \begin{equation}
V_{k+1}\tilde{U}_{k+1}\ket{\eta_{k}}=\frac{\ket 0\ket{\eta_{k+1}}+\ket{e_{k+1}}}{\sqrt{1+\varepsilon_{k+1}}}\mbox{.}\end{equation}
Applying this disentangler and isometry to the whole state of the
chain, one gets\begin{equation}
V_{k+1}\tilde{U}_{k+1}\dots V_{1}\tilde{U}_{1}\ket{\psi}=\frac{\ket 0^{\otimes k+1}\ket{\eta_{k+1}}+\ket{e_{k+1}^{cm}}}{\sqrt{1+\epsilon_{k+1}^{cm}}}\label{eq:chain_state_k+1}\end{equation}
where the accumulated error vector at step $k+1$ is \begin{equation}
\ket{e_{k+1}^{cm}}=\ket{e_{k+1}}+\sqrt{1+\epsilon_{k+1}}V_{k+1}\tilde{U}_{k+1}\ket{e_{k}^{cm}}\label{eq:accumulated_error_vector}\end{equation}
and the square if its norm $\epsilon_{k+1}^{cm}=\braket{e_{k+1}^{cm}}{e_{k+1}^{cm}}$
obeys the recurrence relation $1+\epsilon_{k+1}^{cm}=\left(1+\epsilon_{k+1}\right)\left(1+\epsilon_{k}^{cm}\right)$
since the elementary error vector $\ket{e_{k+1}}$, for which all
previous ancillary particles have been disentangled, is orthogonal
to the vector $V_{k+1}\tilde{U}_{k+1}\ket{e_{k}^{cm}}$. Thus, \begin{equation}
1+\epsilon_{k+1}^{cm}=\prod_{i=1}^{k+1}\left(1+\epsilon_{i}\right)\mbox{.}\end{equation}

\subsubsection{Global error}

After the choice of $m$ disentanglers and $m$ isometries, the reconstructed
state is $\ket{\phi}=V_{m}^{\dagger}\tilde{U}_{m}^{\dagger}\dots V_{1}^{\dagger}\tilde{U}_{1}^{\dagger}\ket 0^{\otimes m+1}\ket{\eta_{m}}$.
Its difference to the actual experimental state $\ket{\psi}$ can
be stated in terms of the (in) fidelity as \begin{eqnarray}
1-\left|\braket{\phi}{\psi}\right|^{2} & = & 1-1/\left(1+\epsilon_{m}^{cm}\right)\nonumber \\
 & = & 1-1/\prod_{i=1}^{m}1+\epsilon_{i}\mbox{.}\label{eq:bound-norm}\end{eqnarray}

Practically, one is interested in guaranteeing that the reconstructed
state is close to the experimental up to global error $E$, i.e.,
to guarantee that $1-\left|\braket{\phi}{\psi}\right|^{2}\leq E$.
Suppose that all error vectors are bounded, i.e. that for all step
$i$, we have $\epsilon_{i}\leq\varepsilon$. Inverting \eqref{eq:bound-norm},
it suffices that \[
\varepsilon\leq(1-E)^{-1/m}-1\simeq E/m\]
in the limit where the tolerable global error $E$ is small. Thus,
we see that \emph{errors accumulate linearly} and that a precision
inversely proportionnal to the number of disentanglers is sufficient
to ensure a constant global error. Furthermore, statistics on the
post-selection performed at each step allows to estimate each $\epsilon_{k}^{cm}$
and in particular $\varepsilon_{m}^{cm}$ that gives direct access
to the distance betweem the reconstructed and experimental states. 

Finally, from this measurement data, one can estimate the error $\epsilon_{i}$
performed at each step in order to identify steps that have gone wrong.
This information can be used to turn the scheme into an \emph{adaptative}
one. Suppose the error is particularly large for a given step. This
might be due to an important amount of entanglement concentrated in
one region of space, \textit{e.g.} near a defect, which can be accounted
for by increasing the MERA refinement parameter $\chi$ locally, i.e.
by using disentanglers acting on a larger number of qubits. In principle,
$\chi$ could be increased until the error is below some threshold.

\subsection{Numerical performance\label{sub:Numerical-performance}}

\subsubsection{Benchmarking results}

We have performed numerical simulations to benchmark the performances
of the MERA learning method. We have generated
random MERA states ---by picking each unitary gate in the circuit
from the unitary group Haar measure---, simulated the experiment on
those states, and use our algorithm to infer the initial MERA state.
We did not introduce noise in measurements to simulate experimental
errors since the error analysis indicates how those errors would build
up. 

As mentioned before, there is no guarantee that our minimization procedure
converges to the true minimum, resulting in small imperfections in
the reconstructed state.%
\begin{figure}[t]
\begin{centering}
\includegraphics[width=1\columnwidth]{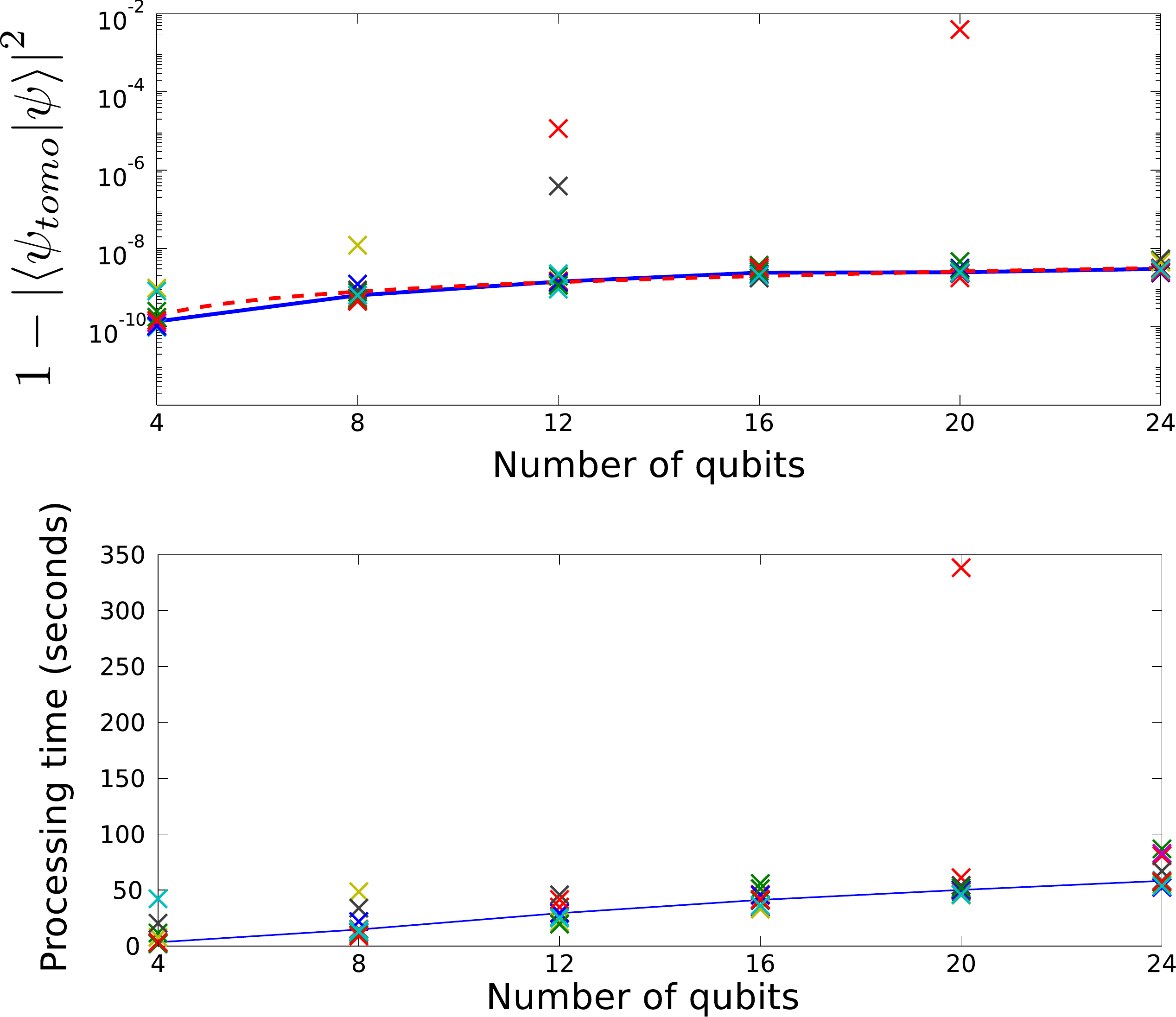} 
\par\end{centering}

\caption{\label{fig:performance-plot}(top) Infidelity to the {}``experimental
state'', \textit{i.e}, $1-\left|\braket{\psi_{tomo}}{\psi}\right|^{2}$
where $\ket{\psi}$ is a random MERA on $n$ qubits and $\ket{\psi_{tomo}}$
is the state reconstructed from the MERA learning method using three
sweeps. (bottom) Processing time (on a standard laptop) to perform
MERA learning using three sweeps. Both figures exhibit 10 runs for
each number of qubits $n\in\{8,\,12,\,16,\,20,\,24\}$. In both figures,
each $\times$ represents results for one random MERA. The full lines
represent median for each number of qubits. The dashed line on the
top figure is the linear approximation to the median. Notice that
the numerical minimisation can fail to converge as illustrated by
the atypical data points. For instance, for one of the 20-qubit MERA,
the processing time was 338.3 seconds and the infidelity to the true
state is large, $1-\left|\braket{\psi_{tomo}}{\psi}\right|^{2}=3.916\times10^{-3}$.}
\end{figure}
 Figure (\ref{fig:performance-plot}, top) shows the distance between
the reconstructed state and the actual state. As indicated by the
dashed line, these results are in good agreement with a linear scaling
of the error where the source of errors is due to finite machine precision
and approximate minimisation of the objective function.

\begin{figure}[t]
\begin{centering}
\includegraphics[width=1\columnwidth]{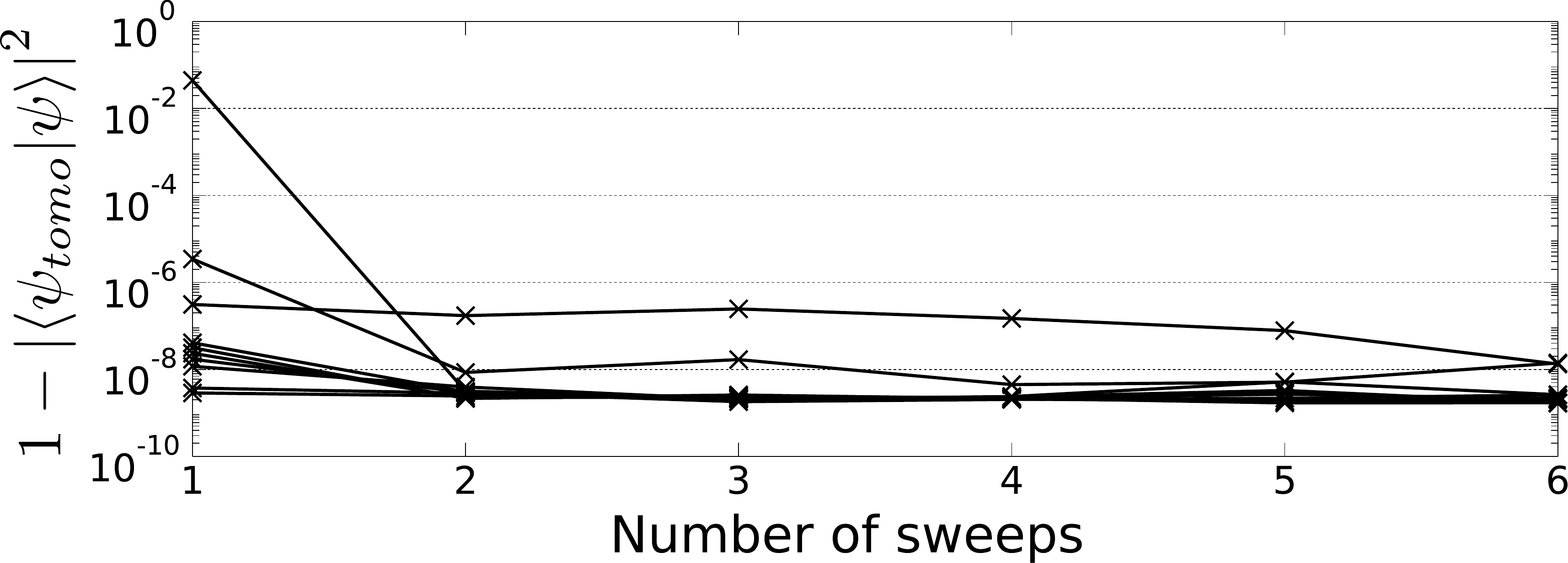} 
\par\end{centering}

\caption{\label{fig:nb_sweeps}Infidelity to a 20 qubit state using a reconstructed
method with a variable number of sweeps. Each line corresponds to
a different random MERA.}
\end{figure}

The inference algorithm's complexity is dominated by the conjugate
gradient descents, and therefore scales linearly with the number of
disentanglers or the number of particles in the system. Figure (\ref{fig:performance-plot},
bottom) shows the actual runtime of the inference algorithm for different
randomly chosen MERA states and of various sizes. Once again, we see
a good agreement with a linear dependence with the system size. Systems
of up to 24 qubits can easily be handled in a few minutes of computation
and requires 1197 different measurement settings for each sweep of
the 24 qubit system. This is to be contrasted with the 656,100 experiments
needed to reconstruct the state of 8 qubits in \cite{Haffner2005}
and the post-processing of the data that took approximately a week
\cite{Blume-Kohout2010}. Additional sweeps allow the method to converge
as showed on Fig. \ref{fig:nb_sweeps}. %

We also tested how our method behaved on a physical model, namely
the 1D Ising model with transverse field at the critical point. The
results obtained where coherent with what is expected from the approximation
of the true ground state with a MERA with refinement parameter $\chi=2$.

\subsubsection{Possible improvements}

Note the presence of isolated points on the graphs of Fig. \ref{fig:performance-plot}
that achieve a lower fidelity and required a longer running time.
These cases appear because the heuristic fails to find a global minimum.
It appears that in some cases, a unitary transformation $U_{23}$
meeting criterion \eqref{eq:Objective-function} is not sufficient
to guarantee that it will be possible to find subsequent disentanglers
obeying \eqref{eq:Objective-function}. Put another way, \emph{locally}
minimising the objective function might not lead to a \emph{global}
optimum. Indeed, consider the following example. Let $\ket{\psi}$
be a MERA state whose first qubit is disentangled from the rest of
the chain, i.e. $\ket{\psi}=\ket 0\ket{\phi}$. The rank of the density
matrix on the first two qubits is at most 2 and that remains true
after \emph{any} unitary is applied on qubits 2 and 3. Thus, any choice
of disentangler minimises the objective function \eqref{eq:Objective-function},
including the identity, \textit{i.e.}, applying no disentangler at
all. However, suppose the state $\ket{\phi}$ on qubits $2$ to $n$
is highly entangled and that removing part of this entanglement between
qubits $2$ and $3$ was crucial to be able to reconstruct its MERA
description. In this case, applying the identity on qubits 2 and 3,
even if locally optimal, was not globally optimal. Hence, minimizing
the objective function \eqref{eq:Objective-function} seems to be
necessary but not sufficient to successively identify all disentanglers.

Although our numerical simulations suggest that this situation is
rather atypical, it is possible to overcome this problem by imposing
additional constraint on the disentangler. For instance, one can demand
that the second qubit be in a state as pure as possible, effectively
minimizing the entanglement between the last qubit of one block and
the first qubit of the next block. This corresponds to the following
modified objective function\begin{equation}
f\left(\tilde{\rho}_{12}\left[U\right]\right)=\sum_{k>2}\lambda_{k}+\epsilon\lambda_{2}\label{eq:modified_objective_function}\end{equation}
\textit{i.e.,} we add a small perturbation that will only take action
when the two smallest eigenvalues of $\tilde{\rho}_{12}[U_{23}]$
are very small and will further constrain the search. This slight
modification solved the problematic situation we considered, and there
exist many other heuristics to improve the method.

\subsection{\label{sub:Trading-unitary-control}Trading unitary control for repeated
measurements}

For pedagogical reasons, we presented our learning method in a way
that required disentanglers and isometries to be physically applied
to the experimental state in order to unravel the circuit. In this
section, we will show how to circumvent unitary control at the price
of slightly more elaborate numerical processing and consuming more
copies of the state. The main idea is to numerically simulate how
measurements performed on the original, unaltered experimental system
would be transformed if the unraveling circuit had been applied.

\subsubsection{Simulating measurements on renormalized state}

\begin{figure}
\begin{centering}
\includegraphics[width=0.9\columnwidth]{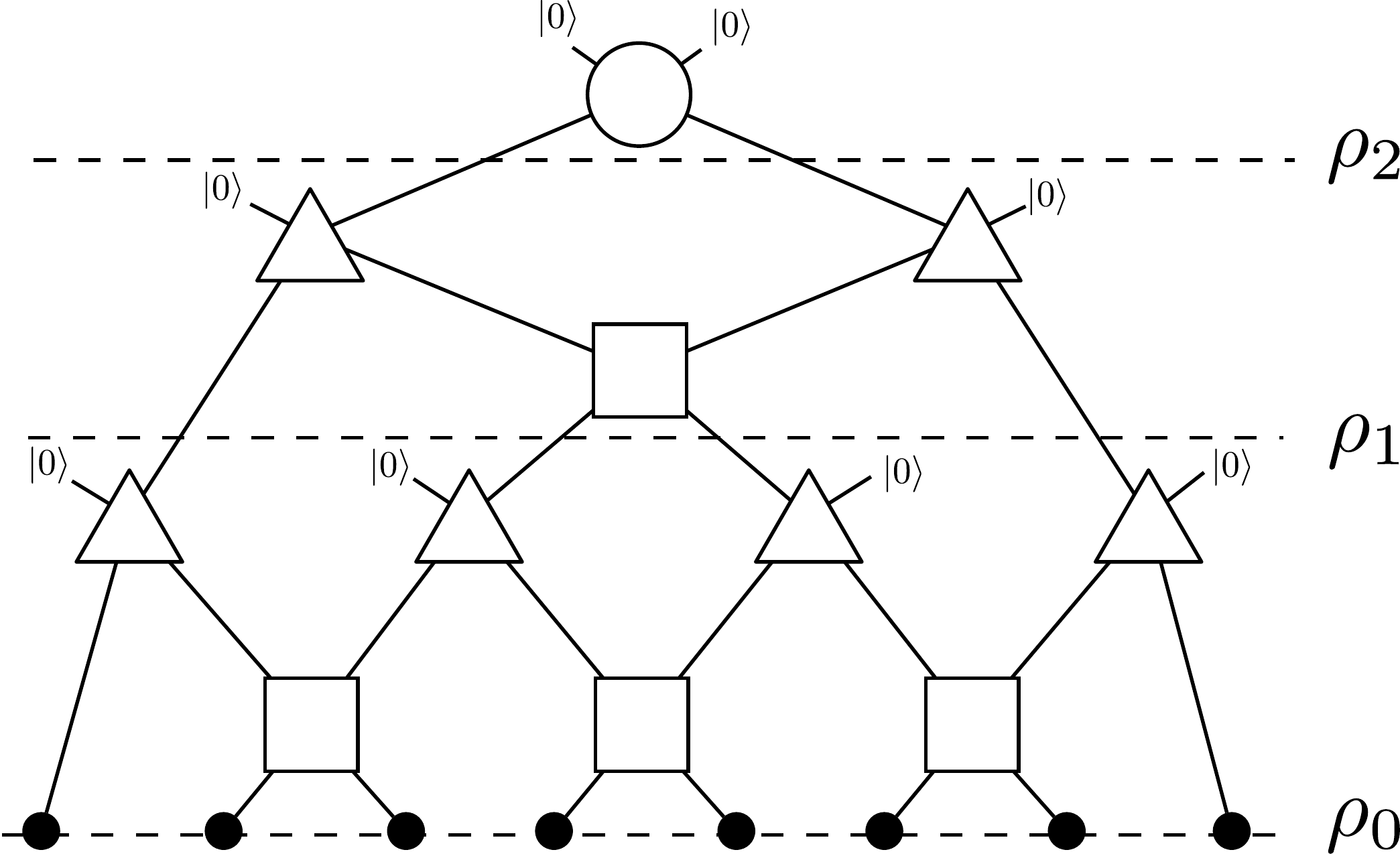}
\par\end{centering}

\caption{\label{fig:MERA-as-renormalisation}MERA as a renormalisation procedure
that creates a sequence of states $\left\{ \rho_{\tau}\right\} _{\tau}$.}
\end{figure}

Recall that a MERA is an ansatz that corresponds to a renormalization
procedure. Each renormalisation step maps a state to another one on
fewer particles and schematically corresponds to a layer of the MERA
circuit. Applying the first layer and removing the ancillary particles
that have been (approximately) disentangled maps the experimental
state $\rho_{0}$ on $n$ particles to a state $\rho_{1}$ on fewer
particles (see Fig \ref{fig:MERA-as-renormalisation}). %
Recursively, this procedure constructs a sequence of states $\left\{ \rho_{\tau}\right\} _{\tau}$.

To get from $\rho_{\tau-1}$ to $\rho_{\tau}$, one can either perform
this mapping \emph{physically} by experimentally applying the gates
corresponding to the MERA layer, or one can \emph{compute} the function
mapping $\rho_{\tau-1}$ to $\rho_{\tau}$ from the description of
the gates. As in \cite{Evenbly2009}, define a ascending superoperator
$\mathcal{A}$ that maps an operator $O_{\tau-1}$ acting on layer
$\tau-1$ to the an operator $O{}_{\tau}$ acting on the next layer
$\tau$\begin{equation}
O_{\tau}=\mathcal{A}_{\tau}(O_{\tau-1})\end{equation}
such that \begin{equation}
\mbox{Tr}[\rho_{\tau}\mathcal{A}_{\tau}(O_{\tau-1})]=\mbox{Tr}[\rho_{\tau-1}O_{\tau-1}]\mbox{.}\end{equation}
This recursively carries over to the experimental state $\rho_{0}$\begin{equation}
\mbox{Tr}[\rho_{\tau}\mathcal{A}_{\tau}\circ\dots\circ\mathcal{A}_{1}(O_{0})]=\mbox{Tr}[\rho_{0}O_{0}]\mbox{.}\end{equation}
Thus, in order to extract information from a density matrix $\rho_{\tau}$,
one can measure the expectation value of several observables $O_{0}^{i}$
on the density matrix $\rho_{0}$. Measuring those observables will
effectively amount to measuring the observables $O_{\tau}^{i}\equiv\mathcal{A}_{\tau}\circ\dots\circ\mathcal{A}_{1}(O_{0}^{i})$
on the density matrix $\rho_{\tau}$. 

\begin{figure}
\begin{centering}
\includegraphics[width=0.9\columnwidth]{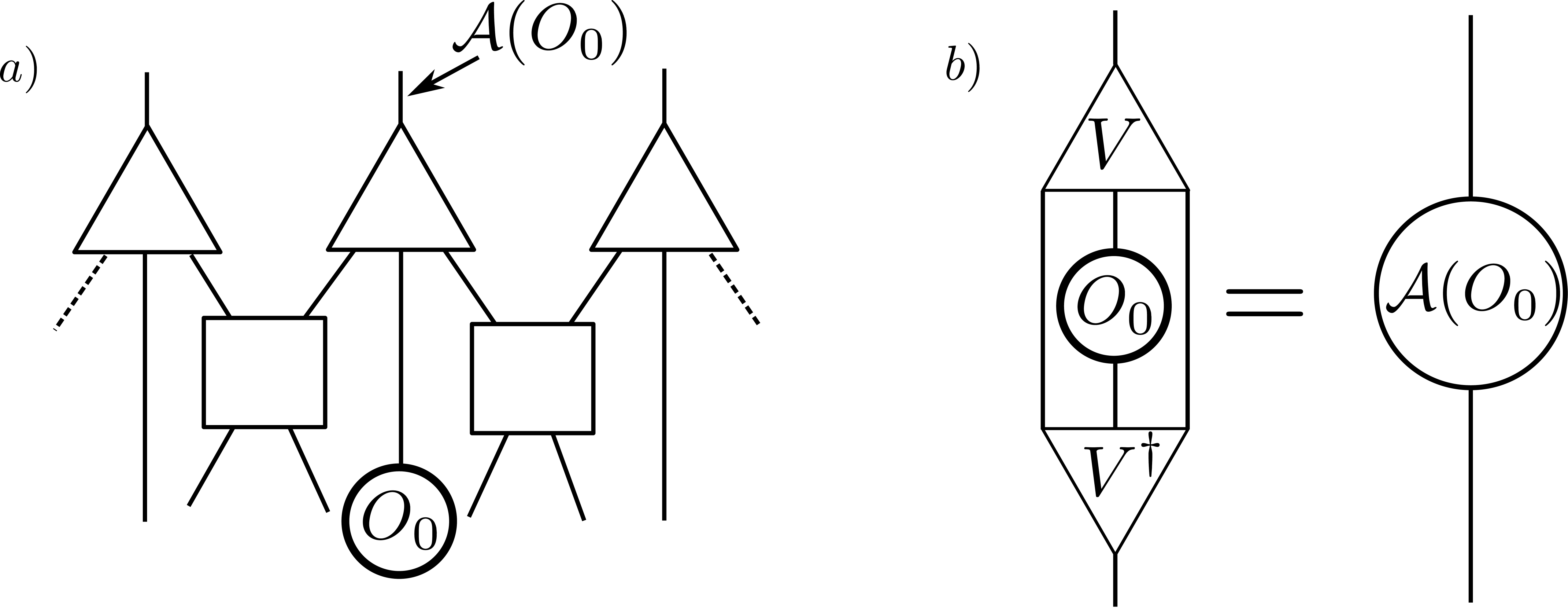}
\par\end{centering}

\caption{\label{fig:Ascending-superoperator}Ascending superoperator and renormalized
observables for a ternary MERA. a) Ternary MERA with one site observable
$O_{0}$ that is transformed into a renormalized observable $\mathcal{A}(O_{0})$
on the renormalized state. b) Tensor contraction corresponding to
the ascending superoperator $\mathcal{A}$.}

\end{figure}

The ascending superoperator can be computed from the knowledge of
the disentanglers and isometries. Its exact form depends on the physical
support of the observable. For instance, for ternary MERA, we can
restrict to ascending superoperator that only depends on the isometries
of the MERA \cite{Pfeifer2009} (see Fig. \ref{fig:Ascending-superoperator}).
 This is a simple example where an experimental observable on one
particle is mapped to observable on one particle. More generally,
observables on many sites will be ascended to observables on fewer
sites. Any choice of observables is valid as long as the renormalized
observables $\left\{ O_{\tau}^{i}\right\} _{i}$ span the support
of the reduced density matrix $\rho_{\tau}$.

\subsubsection{Overhead in the number of measurements}

This procedure leads to a overhead in the total number of measurements
because renormalized observables are less efficient at extracting
information. Suppose (for clarity) that we measure Pauli observables
$\left\{ O_{0}^{i}\right\} _{i}$ on the experimental states. These
observables are orthonormal for the Hilbert-Schmidt inner product
and thus maximize information extraction. However, the renormalized
observables $O_{1}^{i}\equiv\mathcal{A}_{1}(O_{0}^{i})$ need not
be orthonormal. Consider their Gram matrix $G_{ij}=\mbox{Tr}\left[O_{1}^{i}\left(O_{1}^{j}\right)^{\dagger}\right]$
which can be diagonalised by a unitary matrix $Z$. Its normalised
eigenvectors $R_{1}^{i}=\frac{1}{\sqrt{\lambda_{i}}}\sum_{j}Z_{ij}O_{1}^{j}$
are orthornormal observables but cannot be directly measured because
they do not correspond to simple observables on the experimental state,
but instead to linear combination of them. Thus, to reconstruct the
density matrix $\rho_{1}=\sum_{i}r_{1}^{i}R_{1}^{i}$, the expectation
values $r_{1}^{i}=\mbox{Tr}\rho_{1}R_{1}^{i}$ have to be computed
by taking a linear combination of the expectation values $o_{0}^{j}\equiv\mbox{Tr}\rho_{0}O_{0}^{j}$
on the experimental state \begin{equation}
r_{1}^{i}=\frac{1}{\sqrt{\lambda_{i}}}\sum_{j}Z_{ij}\mbox{Tr}\rho_{1}O_{1}^{j}=\frac{1}{\sqrt{\lambda_{i}}}\sum_{j}Z_{ij}o_{0}^{j}\mbox{.}\end{equation}

Due to limited number of repeated measurements, estimation of each
$o_{0}^{i}$ will present a variance $\mathbb{V}(o_{0}^{i})$. Suppose
that measurements are repeated enough times to ensure that all variances
are below a precision threshold, i.e., $\mathbb{V}(o_{0}^{i})\leq\epsilon$.
Since $r_{1}^{i}$ is a linear combination of those measurements,
it will have a variance $\mathbb{V}(r_{1}^{i})=\frac{1}{\lambda_{i}}\sum_{j}|Z_{ij}|^{2}\mathbb{V}(o_{0}^{j})\leq\epsilon/\lambda_{i}$.
Therefore, in order to ensure a precision $\epsilon$ on the estimate
of $r_{1}^{i}$, this imprecision needs to be compensated by multiplying
the number of repeated measurements by the \emph{conditioning factor}$\lambda_{i}^{-1}$. 

When scaling operators on $\tau$ layers, the conditioning factors
for each layer will multiply (in the worst case) but we expect the
conditioning for each layer to be a constant independant of system
size. Thus, the total number of measurements will remain \emph{polynomial}
in the number of particles since there is only a \emph{logarithmic}
number of renormalisation layers. 

We can make this argument rigorous for critical systems that exhibit
scale-invariance, precisely the physical systems for which MERA was
introduced. Due to scale-invariance, the ascending operator $\mathcal{A}_{\tau}$
will not depend on the index of the layer and we refer to it as the
scaling superoperator $\mathcal{S}$ \cite{Pfeifer2009}. Its diagonalization
yields eigenvectors $\phi_{\alpha}$ called scaling operators associated
to eigenvalues $\mu_{\alpha}$. In \cite{Pfeifer2009}, it was shown
that those eigenvalues are related to the scaling dimensions $\Delta_{\alpha}$
of the underlying conformal field theory (CFT) by $\Delta_{\alpha}=\log_{3}\mu_{\alpha}$
where the basis of the log depends on the MERA type (here we consider
a ternary MERA for clarity). Scaling operators $\phi_{\alpha}$ can
be used as observables to extract information about states in higher
level of the MERA. Indeed, one can simulate a measurement of $\mathcal{S}^{\tau}(\phi_{\alpha})$
on $\rho_{\tau}$ by measuring the observable $\phi_{\alpha}$ on
$\rho_{0}$. We can analyze the increase in the number of measurements
by distinguishing two sources of imprecision. First, to reconstruct
$\rho_{\tau}$ one has to use normalised operator $\phi_{\alpha}^{[\tau]}=3^{\tau\Delta_{\alpha}}\mathcal{S}^{\tau}(\phi_{\alpha})$
whose increased statistical fluctuations have to be compensated by
performing additional measurements. Second, diagonalising the Gram
matrix of the $\phi_{\alpha}^{[\tau]}$ will introduce another conditioning
factor. However, this Gram matrix is independant of the layer since
$G_{\alpha\beta}^{[\tau]}=\mbox{Tr}\left[\phi_{\alpha}^{[\tau]}\phi_{\beta}^{[\tau]}\right]=\mbox{Tr}\left[\phi_{\alpha}\phi_{\beta}\right]$.
Thus, the conditioning factor for layer $\tau$ will be the product
of a factor exponential in the number of layers and a constant factor
coming from the orthonormalisation.

This modified scheme circumvents the need of unitary control, but
looses some of the features of the original scheme. First, because
the system is not physically disentangled, we cannot certify directly
the fidelity of the reconstruction. Second, as explained in appendix
\ref{sec:Error-analysis-no-post-select}, the errors build up quadratically.

\section{\label{sec:Discussion}Discussion}

In this work, we have presented a tomography method that allows to
efficiently learn the MERA description of a state by patching together
tomography experiments on a few particles and using fast numerical
processing. The method is heuristic but works very well in numerical
simulations. A complete analytical understanding of how to find an
optimal disentangler at each step would be desirable, but may well
be intractable. With regards to experimental use, the method should
be tought of as a proof of principle and is flexible enough to be
adapted to accomodate many experimental constraints.

One issue of fundamental interest raised by our work is the relationship
between the numerical tractability of a variational class of states
and the ability to learn efficiently the variational parameters. In
order to be interesting, variational class of states must not only
be described by a small number of parameters, but also allow for the
efficient numerical computation of quantities of interest, such as
the energy of the system, correlation functions, or more generally
expectation values of local observables. On its own, an efficient
representation is of limited computational usefulness. For instance,
the Gibbs state or thermal state of a local Hamiltonian is described
by a few parameters --- a temperature and a local Hamiltonian ---
but does not allow to extract physical quantities of interest efficiently.
Another example is the variational class of projected entangled pair
states or PEPS \cite{VMC08}, the generalization of MPS to system
in more than one dimension. While PEPS have been instrumental in better
understanding of quantum many-body systems, they are in general intractable
numerically \cite{SWV+07}. 

Is there a relation between numerical tractability and efficient tomography?
The method presented in \cite{CPF+10} to learn a MPS from local measurements
made explicit use of the energy minimization algorithm for MPS; namely
DMRG \cite{Schollwock05}. This example suggests that numerical tractability
could imply that learning the variational parameters is possible.
In that regard, MERA are intriguing states because they live at the
frontier of tractability. Indeed, in more than 1 dimension, MERA states
are a subclass of PEPS \cite{VC04} with a bond dimension independant
of system size \cite{BKE10}. While the computation of expectation
values of local observables is believed to be intractable for PEPS,
it is efficient for MERA. In one dimension, MERA can be seen as MPS
if one allows the bond dimension to grow \emph{polynomially} with
the size of the system (while MPS are usually required to have a \emph{constant}
bond dimension). Thus, while MPS manipulations typically have a computational
cost linear in the number of particles, 1D-MERA manipulations have
a computational cost which is superlinear (but yet polynomial).

Beyond MPS and MERA, one could consider states obtained from a quantum
circuit where the positions of the gates are known and try to identify
those gates. An interesting question is then to characterize what
topology of circuits makes it possible to learn gates efficiently.
This could lead to formal methods for the testing and verification
of quantum hardware. 
\begin{acknowledgments}
OLC acknowledges the support of the Natural Sciences and Engineering
Research Council of Canada (NSERC) through a Vanier scholarship. OLC
wishes to thank Andy Ferris for stimulating discussions and sharing
numerical data on scale-invariant MERAs. DP acknowledges financial
support by the Lockheed Martin Corporation.
\end{acknowledgments}
\bibliographystyle{apsrev}

\bibliographystyle{apsrev}
\bibliography{ref}

\appendix

\section{\label{sec:Error-analysis-no-post-select}Error analysis without
post-selection}

The modified scheme that circumvents the need of unitary control modifies
the error propagation. Indeed, the scaling of the overall error increases
since the error at each step will depend on previous errors. Essentially,
to find the optimal disentangler, the algorithm will not receive the
perfect state $\ket{\eta_{k}}$ (see eq. \eqref{eq:state-with-error})
but the state $\frac{1}{1+E_{k}^{cm}}\proj{\eta_{k}}+\frac{E_{k}^{cm}}{1+E_{k}^{cm}}\proj{E_{k}^{cm}}$
where $\ket{E_{k}^{cm}}$ is a subnormalised error vector resulting
from the accumulation of all previous errors whose square norm is
$E_{k}^{cm}\equiv\braket{E_{k}^{cm}}{E_{k}^{cm}}$. Thus, the numerical
minimisation returns a unitary that is not the perfect disentangler.

In the degenerate case ---when there are many unitaries reaching roughly the
same minimum value of the objective function---, this might change the
disentangling unitary
drastically. Indeed, we note that the existence of degenerate local minima
affects both of our tomography methods, the one with and the one without unitary
control of the system. In such degenerate cases, exploring many local minima by
selecting random initial guesses could get around the problem. However, it is
likely that these instances are intrinsically hard and that our algorithm does
not converge to the right answer in those cases, c.f. the atypical data points
on Fig. \ref{fig:performance-plot}. 

In the
non-degenerate case, we can argue that the accumulation of errors
would be \textit{quadratic} in the number of particles. We proceed
in three steps. First, we analyze how the modification of the input
state will affect the disentangling unitary returned by the algorithm.
Second, we evaluate how this imperfect disentangler impacts the error
propagation. Third, we bound the error to show the quadratic scaling.

Our algorithm returns the unitary $\tilde{U}=e^{i\tilde{H}}$ that
minimizes the objective function \eqref{eq:Objective-function} for
a given state $\rho$. If we don't post-select on the ancillary particles
being disentangled, this minimization is not performed on the the
perfect state $\rho_{0}=\ketbra{\eta_{k}}{\eta_{k}}$ but rather on
the state $(1-\varepsilon)\rho_{0}+\varepsilon\sigma$ where $\varepsilon=\frac{E_{k}^{cm}}{1+E_{k}^{cm}}$
and $\sigma=\proj{E_{k}^{cm}}$. We want to know how much $\tilde{U}=\arg\min_{U}f(U,\rho)$
changes when $\rho$ changes. Using the chain rule, we formally write
$\frac{\partial\tilde{U}}{\partial\rho}=\frac{\partial\tilde{U}}{\partial f}\frac{\partial f}{\partial\rho}$.
The first term quantifies how much $\tilde{U}$ changes when the objective
function changes for a given $\rho$. In the non-degenerate case,
we expect this term to be bounded in norm by a Lipschitz constant
$\eta$. The second term evaluates how the objective function changes
when going from $\rho_{0}$ to $(1-\varepsilon)\rho_{0}+\varepsilon\sigma$.
Recalling that the objective function is a sum of eigenvalues and
using non-degenerate perturbation theory, this term is going to be
proportional to $\varepsilon$. Thus, instead of $\tilde{U}=e^{i\tilde{H}}$,
the minimization algorithm returns $e^{i(\tilde{H}+\varepsilon\eta A)}\approx W\tilde{U}$
where $W=e^{i\varepsilon\eta A}$.

As a consequence, eq. \eqref{eq:chain_state_k+1} is modified to read

\begin{equation}
W_{k+1}\frac{\ket 0^{\otimes k+1}\ket{\eta_{k+1}}+\ket{e_{k+1}^{cm}}}{\sqrt{1+\epsilon_{k+1}^{cm}}}=e^{i\theta}\frac{\ket 0^{\otimes k+1}\ket{\eta_{k+1}}+\ket{E_{k+1}^{cm}}}{\sqrt{1+E_{k+1}^{cm}}}\mbox{.}\end{equation}
Taking into account the anomalous unitary $W$, we get \begin{equation}
1+E_{k+1}^{cm}=1+\epsilon_{k+1}^{cm}/\beta^{2}=(1+\epsilon_{k+1})(1+E_{k}^{cm})/\beta^{2}\end{equation}
where $\beta^{2}=\left|\bra{\eta_{k+1}}\bra 0^{\otimes k+1}W\ket 0^{\otimes k+1}\ket{\eta_{k+1}}\right|^{2}=\left|\langle W\rangle\right|^{2}$.
Using $W=e^{i\varepsilon\eta A}$, calculations show that \begin{equation}
\beta^{2}=1-\varepsilon^{2}\eta^{2}\left(\langle A^{2}\rangle-\langle A\rangle^{2}\right)=1-\varepsilon^{2}\eta^{2}\Delta^{2}\end{equation}
where the variance $\Delta^{2}$ of $A$ with respect to state $\ket 0^{\otimes k+1}\ket{\eta_{k+1}}$
appears. Recalling that $\varepsilon=\frac{E_{k}^{cm}}{1+E_{k}^{cm}}$,
one gets\begin{equation}
\beta^{2}=\frac{\left(1+E_{k}^{cm}\right)^{2}-\left(E_{k}^{cm}\right)^{2}\eta^{2}\Delta^{2}}{\left(1+E_{k}^{cm}\right)^{2}}\geq\frac{1}{\left(1+E_{k}^{cm}\right)^{2}}\end{equation}
for any $E_{k}^{cm}$ if $\eta^{2}\Delta^{2}\leq1$ or for small $E_{k}^{cm}$
otherwise.

Thus, the error magnitude $E_{k+1}^{cm}$ obeys the recurrence relation\begin{eqnarray*}
1+E_{k+1}^{cm} & \leq & \left(1+E_{k}^{cm}\right)^{3}(1+\epsilon_{k+1})\\
 & \leq & \left(1+\epsilon_{1}\right)^{3k}\dots\left(1+\epsilon_{k+1}\right)\mbox{.}\end{eqnarray*}
Assuming that the error at each step is bounded $\epsilon_{k}\leq\epsilon$,
the total error scales as 

\begin{equation}
E_{m}^{cm}\leq\frac{3}{2}m^{2}\epsilon\mbox{.}\end{equation}

\section{\label{sec:Contraction}Comparing a reconstructed MERA to a predicted
MERA}

In this section, we describe a polynomial algorithm to contract two
MERA states, thus allowing to compute their fidelity. This algorithm
is of practical interest for comparing a MERA whose parameters have
been identified experimentally using our method to a predicted MERA
state --found by numerical optimisation for instance. Notice that
contracting two different MERA states also allows to compute expectation
values of tensor product of local observables $\bigotimes_{i}A_{i}$
since it suffices to contract the original state $\ket{\psi}$ and
the modified state $\ket{\phi}=\bigotimes_{i}A_{i}\ket{\psi}$, which
is also a MERA state. 

Defining a method to contract two MERA states is equivalent to giving
a prescription on how to sequentially contract the tensor network
resulting from joining two MERA states. Recall that contracting two
tensors $(M)_{i_{a}j_{b}}$ and $(N)_{k_{b}\ell_{c}}$ to obtain $T_{i_{a}\ell_{c}}=\sum_{j_{b}}M_{i_{a}j_{b}}N_{j_{b}\ell_{c}}$
has a computational cost of $a\times b\times c$ where $a$ is the
number of values that the index $i_{a}$ can take $b$ and $c$ are
defined in the same way with respect to $j_{b}$ and $\ell_{c}$.
In a tensor network, every tensor is usually represented with a number
of bonds that each represent an index that has the same maximal number
of possible values. For a MERA, this maximal bond dimension is usually
denoted by $\chi$. 

\begin{figure}
\begin{centering}
\includegraphics[width=1\columnwidth]{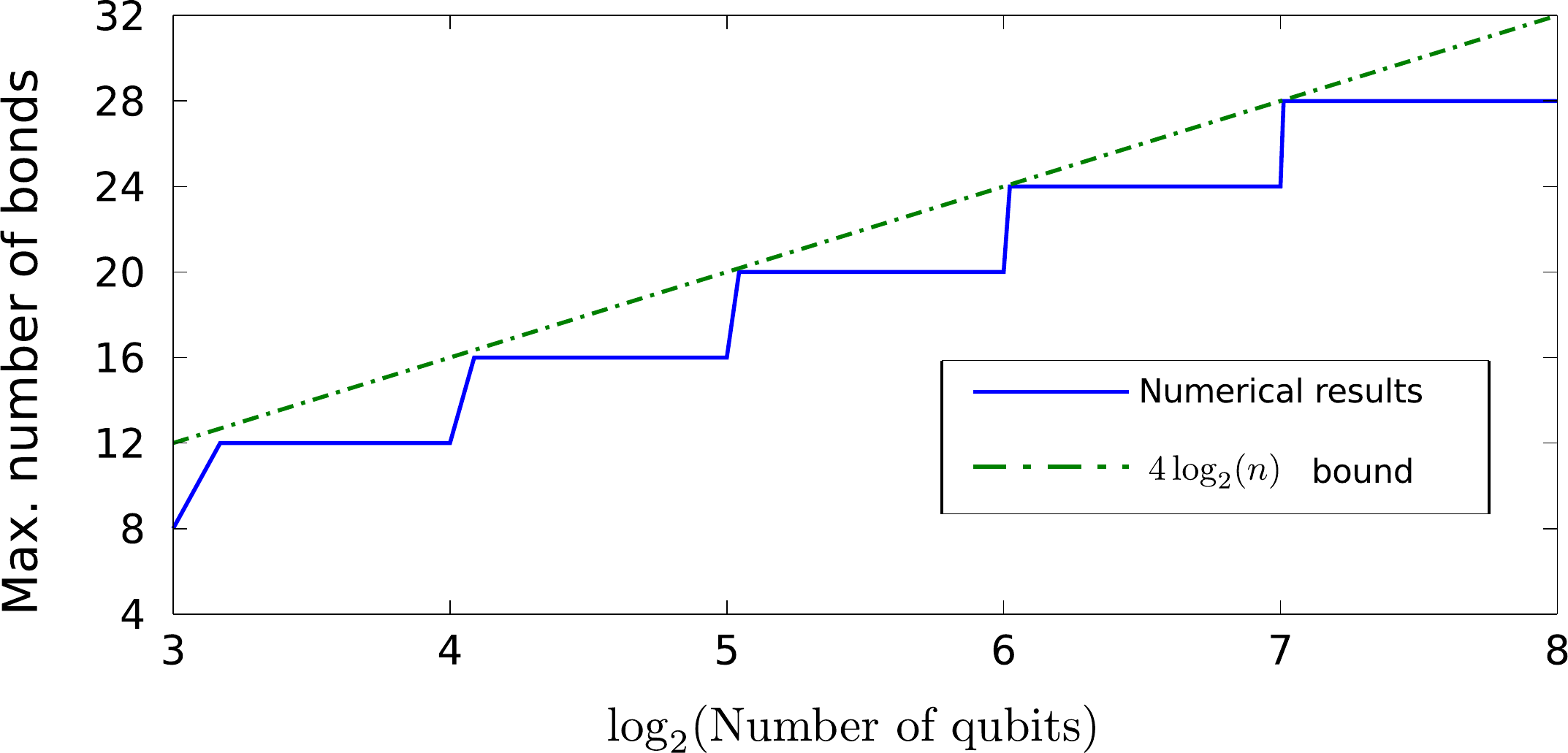}
\par\end{centering}

\begin{centering}
\includegraphics[width=1\columnwidth]{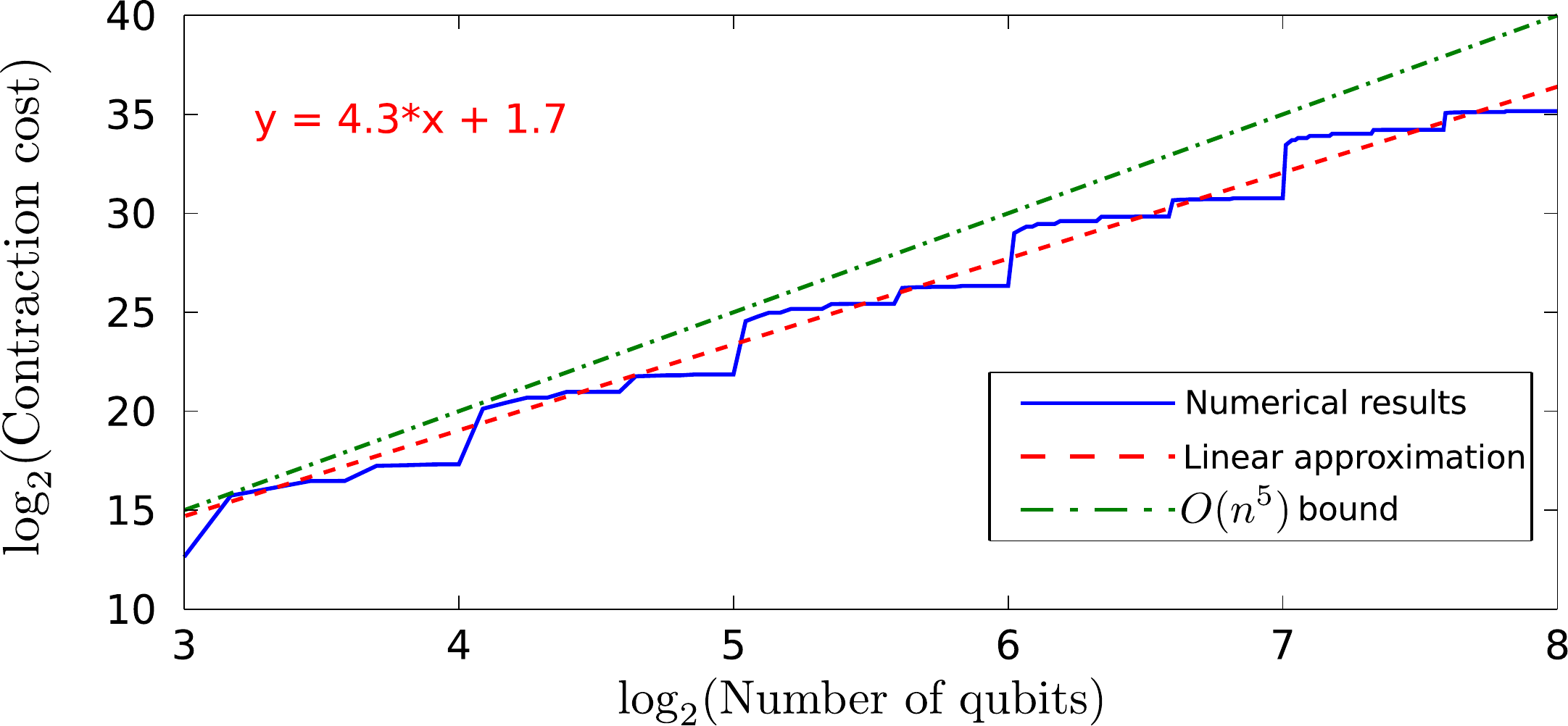} 
\par\end{centering}

\caption{\label{fig:contraction}(top) Maximum number of bonds during the contraction
procedure as a function of the logarithm of the number of qubits $n$.
Numerical results (solid blue line) are consistent with the expected
bound of $4\log_{2}n$. (bottom) Contraction cost $C$ as a function
of the number $n$ of qubits on a logscale. Numerical results (solid
blue line) are consistent with the $O(n^{5})$ bound (dot dashed green
line) but linear approximation (dashed red line) indicate that the
cost scales like a smaller power of $n$, namely $C\simeq n^{4.3}$.}
\end{figure}

The main idea to contract efficiently two MERA states is essentially
to turn them into two MPS before contracting them. We look at the
MERA circuit as having $n/2$ columns of gates vertically and $\log_{\chi}n-1$
renormalisation layers horizontally. The sequence of contraction is
to sequentially contract every tensor in the leftmost column to create
a tensor with a large number of bonds that will then contract with
every tensor in the next column. The maximal number of bonds that
this leftmost tensor will have throughout the contraction of the network
is given by the maximal number of bonds that are opened when taking
a vertical cut in the tensor network. For a single MERA, cutting through
each of the $\log_{\chi}n-1$ layer opens up two bonds, one for the
righmost incoming edge of the isometry and one for the outgoing edge
of the isometry. Thus, for the contraction of two MERAs, the maximum
number of bonds for a vertical cut is bounded by $\max\#=2\times2\times\log_{\chi}n=4\log_{\chi}n$,
which is verified numerically (see top of Fig. \ref{fig:contraction}).
 Since at every contraction step, the leftmost tensor with a large
number of bonds contract with another tensor that has at most two
bonds in addition to the ones being contracted, the maximum cost of
one contraction is $\chi^{\max\#}\chi^{2}=\chi^{2}n^{4}$. Finally,
there are $O(n)$ disentanglers and isometries to contract so the
total cost of contracting the network is bounded by $O(n^{5})$. Actual
numerical simulations show that this bound is probably not tight (see
bottom of Fig. \ref{fig:contraction}).
\end{document}